# Rotational and Irrotational wind forcing as dual drivers of El Niño-Southern Oscillation variability

by

Gian Luca Eusebi Borzelli[1*], Cosimo Enrico Carniel[2], Sandro Carniel[3], Mauro Sclavo[3]

[1]Center for Remote Sensing of the Earth (CeRSE), Rome Italy. e-mail: luca_borzelli@yahoo.it; gianluca.borzelli@cerse.it

[2]Institute for Atmospheric and Climate Sciences, ETH Zürich, Zurich, Switzerland.

[3]CNR, Istituto di Scienze Polari, Venezia Italy.

*Corresponding author

**Abstract.**

El Niño–Southern Oscillation (ENSO) is the Earth's strongest source of interannual climate variability. Although its center of action is in the tropical Pacific, it has significant influences on the climate at the planetary scale. ENSO is sustained by a feedback process between equatorial winds, vertical displacements of the thermocline, and sea surface temperature (SST) anomaly gradients. This produces an oscillation in the SST anomaly between a warm (El Niño) and a cold phase (La Niña). While the natural time-scale of ENSO variability is interannual, variations in its behavior and characteristics have been observed over longer time scales, including decadal and interdecadal. To account for changes in the action of the wind stress over the slowly changing stratification of the ocean, we introduce a dimensionless wind stress (DWS) and use the Helmholtz decomposition to break it down into two components: irrotational (curl-free) and solenoidal (divergence-free). We show that the irrotational component of the DWS drives the dynamics of the thermocline on the interannual time-scale, while the solenoidal component, whose action depends on the stratification of the water column, affects the variability of the thermocline over decadal or interdecadal time-scales. Furthermore, we develop an integral relation that links the variability of the thermocline depth anomaly across the entire tropical Pacific to the variability of the average SST anomaly in the central tropical Pacific (Niño-3.4 index) and conclude that the DWS irrotational component determines the interannual variability of the Niño-3.4 index, while the solenoidal component determines its long-term variability.

1. **Introduction**

Since the 17th century, Peruvian fishermen have observed a recurring anomalously warm current along the Peruvian coast near Christmas. They named this phenomenon "El Niño" (the little boy) or "Niño de Cristo" (Christ child). The first scientific mention of "El Niño" dates back to the early 1890s (Carrillo, 1892). However, until the first half of the 20th century, it was believed to be a local occurrence limited to the coasts of Peru and Ecuador. Since then, significant discoveries have been made about the structure and time-space scales involved in El Niño. It is now understood to be a manifestation of the strongest year-to-year climate fluctuation on our planet known as El Niño Southern Oscillation (ENSO), a natural oscillation of the tropical Pacific climate system that is sustained by a feedback process between equatorial trade winds, vertical displacements of the thermocline, and sea surface temperature (SST) gradients (e.g., Bjerknes, 1969; Wyrtki, 1975). This results in an irregular oscillation between warm (El Niño) and cold (La Niña) phases, with peaks occurring in boreal winter and recurring every 2-5 years (Rasmusson et al., 1990; Jiang et al., 1995).

Our understanding of the oscillatory nature of ENSO relies on recognizing the relationship between changes in internal ocean dynamics and changes in the wind stress pattern. By interpreting the tropical Pacific ocean as a double-layer system, the heat stored in the water column depends on the depth of the interface that separates the surface and bottom layers. The zonal deformation of this interface (i.e. the dynamical ocean subsurface memory) is sustained by the dynamical balance between the fluid internal zonal pressure gradient and zonal winds (Cane and Zebiak, 1985; Philander, 1990; Jin, 1997; Burgers et al., 2005; Jin et al., 2006).

Therefore, the SST anomaly, depth of the interface separating the surface and bottom layers, whose fluctuations from a reference depth are often assumed to coincide with fluctuations in the depth of a sentinel isotherm (typically the 15 ºC, 17 ºC, or 20 ºC isotherm depth referred to as $Z_{15}$, $Z_{17}$, $Z_{20}$), and zonal wind stress are a natural set of "state variables" that can be used to infer the basic physics underlying ENSO. In a broad sense, and assuming that over the interannual time-scale the tropical Pacific climate system can be approximated by a linear system, the dynamics underlying ENSO is described by linear models that relate the evolution of the state variables to each other (Penland and Sardeshmukh, 1995; Newman et al., 2009; Newman et al., 2011; Capotondi and Sardeshmukh, 2015).

Over the past 25 years, the linear recharge/discharge oscillator (LRDO) model, proposed by Jin (1997), Burgers et al. (2005), and Jin et al. (2006), has been a reliable conceptual benchmark for understanding the interannual variability of SST anomalies in the tropical Pacific, despite some discrepancies with observations. This model describes the ocean as a double-layer system, in which the zonal deformation of the interface between the surface and bottom layer is in equilibrium with a purely zonal wind stress. In this representation, the equilibrium between zonal wind stress and internal oceanic zonal pressure gradients is achieved through a linear relationship between zonal wind stress in the central tropical Pacific and the difference in depth of the interface layer between the eastern and western parts of the basin. The oscillation is initiated by the action of internal downwelling Kelvin waves (KWs), which transfer the heat eastward and give rise to an oscillation of the thermocline. Picaut and Delcroix (1995) focused on equatorial wave sequences associated with the 1986-1987 El Niño and showed that anomalous zonal advection was linked to KWs excited by a succession of local westerly winds (WWs) in the western/central tropical Pacific. This highlights the central role of WWs in triggering SST anomalies over the interannual time-scale (see also Wang and Fiedler, 2006; Santoso et al., 2017).

In a more recent study, Eusebi Borzelli and Carniel (2023) discussed the zonal displacements of the convergence region between WWs and easterly winds (EWs). The authors showed that internal KWs are regularly formed in the tropical Pacific beneath the WW/EW convergence and that the zonal displacements of the WW/EW convergence were highly correlated with the Southern Oscillation Index (SOI). In the terminology used by the authors, the WW/EW convergence was the region of minimum wind stress divergence (i.e. maximum of the wind stress divergence absolute value), highlighting the crucial role of wind stress divergence in determining SST anomaly variability over the interannual time-scale.

Over the long term, time series of ENSO indexes exhibit a ''regime shift'' from 1976, first noted by Quinn and Neal (1984, 1985). Subsequent studies (Trenberth, 1990; Trenberth and Hurrell, 1994; Graham, 1994), have emphasized the importance of this feature in relation to interdecadal climate variability over the North Pacific. Since then, several studies have concentrated on the interdecadal or longer ENSO variability. In this context, it is important to note that the behavior and characteristics of ENSO are closely linked to the slowly changing background mean climate conditions of the equatorial Pacific. These conditions impact the feedback processes of ENSO (Fedorov and Philander, 2001; An et al., 2008; Collins et al., 2010). It is projected that the mean climate state of the tropical Pacific will change due to global warming, with a faster rate of warming in the equatorial Pacific compared to off-equatorial regions (Liu et al., 2005; Collins et al., 2010). This warming is expected to be more pronounced in the eastern and western equatorial Pacific compared to the central region (Xie et al., 2010). Additionally, the surface of the ocean is expected to warm more quickly than the subsurface (An et al., 2008). This surface warming pattern is responsible for an increase in rainfall, particularly in the eastern part of the basin, which is a defining characteristic of extreme El Niño events (Zhong et al.,

2025). Accordingly, Cai et al. (2015; 2022) observed that El Niño events are becoming more frequent and intense under global warming.

From a slightly different perspective, some research has focused on the effect of meridional mean-state change on low frequency variability of El Niño. It is well known that the intertropical convergence zone (ITCZ) is typically located north of the equator for most of the year, resulting in climatological cross-equatorial southerly winds across the eastern equatorial Pacific. The relaxation of these southerly winds crossing the equator leads to a weakening of the cold and westward currents south of the equator, contributing to the thermocline downwelling and SST warming in the eastern equatorial Pacific, which in turn modulates El Niño intensity (Philander and Pacanowski 1981; Périgaud et al. 1997; Hu and Fedorov, 2018; Peng et al., 2020). This observation has led some authors to investigate ENSO variability over interdecadal or longer time scales in relation to low frequency variability of the ITCZ and meridional cross-equatorial southerly winds (Xie et al., 2018; Capotondi and Ricciardulli, 2021; Wang and Zheng, 2023).

Here, assuming the tropical Pacific as a double-layer ocean, we use changes in the $Z_{15}$ as a proxy for fluctuations in the depth of the interface separating the surface and bottom layers, to develop an integral relationship that links changes in the $Z_{15}$ anomaly across the entire tropical Pacific with the Niño-3.4 index ($N_{34}$, i.e. the five-month median filtered average SST anomaly over the region 5ºS-5ºN, 170ºW-120ºW, termed to as Niño-3.4 region). To account for changes in the wind stress's effect on the slowly changing stratification of the ocean, we introduce a dimensionless wind stress (DWS). The DWS is defined as the ratio of the wind stress to the product of the surface-layer density and the square of the first internal mode velocity, with the first internal mode velocity defined as $c_1=(g'H)^{1/2}$ with $H$ representing the depth of the surface layer and $g'$ the reduced gravity (see appendix B for details). We then use the Helmholtz decomposition to break the DWS down into two components: irrotational (curl-free) and solenoidal (divergence-free). Our analysis shows that interannual variations in the $N_{34}$ can be explained by the coupling between the irrotational part of the DWS and fluctuations in the thermocline depth. Furthermore, we observe that uneven temperature changes in the surface and bottom layers of the ocean (An et al., 2008) lead to changes in the reduced gravity field ($g'$), which in turn affects the mean dynamics of the ocean. We demonstrate that the coupling between the solenoidal component of the DWS and slow changes in the ocean dynamics can explain interdecadal variability of the $N_{34}$, including changes in the intensity of El Niño events.

## 2. The data

Surface winds and SST were obtained from the Copernicus data archive. Copernicus provides parameters over a regular spatial grids of 0.25°×0.25°. This study utilized ERA5 monthly surface zonal winds and SST over the region 160°E-90°W, 5°S-5°N in the period Jan 1960-Dec 2020.

The wind stress was calculated from surface wind components using the standard formula $\tau=\rho_{air} \cdot c_D \cdot (v_x^2+v_y^2)^{1/2} \cdot v$, where $v=(v_x, v_y)$ is the surface wind, $\rho_{air}$ is the air density at the ocean surface (taken as 1.2 kg/m³) and $c_D$ is the drag coefficient (taken as $1.2 \cdot 10^{-3}$ if $(v_x^2+v_y^2)^{1/2}<11$ m/s and $[0.49+0.065 \cdot (v_x^2+v_y^2)^{1/2}] \cdot 10^{-3}$ if $(v_x^2+v_y^2)^{1/2} \geq 11$ m/s).

SST data were used to compute the $N_{34}$. Monthly SST data from January 1960 to December 2020, were extracted from the Copernicus database over the Niño-3.4 region. The SST anomaly was computed using as a reference climatological baseline, the period January 1991-December 2020.

The Institute of Atmospheric Physics (IAP) at the Chinese Academy of Sciences (CAS) has provided salinity and temperature (TS) data for the water column. These data consist of monthly objective maps covering the entire global ocean from January 1960 to December 2020, sampled over a regular spatial grid of 0.5°×0.5°. The data includes TS profiles from the surface to a depth of 2000 m (for more information, refer to Cheng et al., 2017). In this study, data from the region (160°E-90°W, 5°S-5°N) were extracted to compute seawater density profiles.

## 3. The $Z_{15}$-$N_{34}$ integral relationship

Our objective is to explain the variability of $N_{34}$ using the $Z_{15}$ anomaly through an integral relation of the form

$$N_{34}(t)=\int_{(S)} \frac{dx}{S} w(x) Z(x,t) = \sum_{m=1}^{M} Z(x_m,t) \cdot w(x_m) \qquad (1)$$

In this equation $Z(x_m,t)$ represents the $Z_{15}$ anomaly at point $x_i$ on date $t$, and $N_{34}(t)$ represents the $N_{34}$ on date $t$. The unknowns $w(x_m)$ define the linear model that links $Z$ and $N_{34}$. We determine $w(x_m)$ by requiring that the representation (1) is "optimal" in the least-square sense, namely we require that the Root Mean Square (RMS) between $N_{34}$ and its representation on the right-hand side of (1) is minimum, i.e.

$$min\left\{\left\langle\left[N_{34}(t)-\sum_{m=1}^{M} Z(x_m;t) \cdot w(x_m)\right]^2\right\rangle\right\}_{w(x_1),w(x_2),...,w(x_M)} \qquad (2)$$

where the bra-ket symbol denotes temporal averaging. By computing the derivative with respect to $w(x_k)$ of the RMS, and requiring the minimum condition, we obtain a system of linear equations for the unknowns $w(x_m)$ ($m=1...M$), i.e.

$$\sum_{m=1}^{M} A(x_k, x_m) \cdot w(x_m) = n(x_k) \tag{3}$$

where

$$n(x_k) = \int_0^T \frac{dt}{T} N_{34}(t) \cdot Z(x_k, t)$$
$$A(x_k, x_m) = \int_0^T \frac{dt}{T} Z(x_k, t) \cdot Z(x_m, t) \tag{4}$$

The matrix $A$, with elements $(A)_{km} = A(x_k, x_m)$, is the autocorrelation matrix of $Z$. Taking this in to account, denoting with $\lambda_n$ the variance explained by $n^{th}$ Empirical Orthogonal Function (EOF) of $Z$, with $Y_n$ the $n^{th}$ dimensionless EOF, and with $\alpha_n$ its corresponding dimensionless Principal Component (PC), we have (see Appendix A for the calculation)

$$w(x_m) = \sum_n \frac{c_n}{\sqrt{\lambda_n}} \cdot Y_n(x_m) \tag{5}$$

where the sum is extended to all EOFs of $Z$, and $c_n$ is the projection of the $N_{34}$ along the direction of the $n^{th}$ PC, i.e.

$$c_m = \int_o^T \frac{dt}{T} \alpha_m(t) \cdot N_{34}(t) \tag{6}$$

Note that $w(x)$ is a transfer function that maps the $Z_{15}$ anomaly over the entire tropical Pacific to the average SST anomaly over the central tropical Pacific. In Figure 1a, we show the measured $N_{34}$ (blue solid line) and the $N_{34}$ rebuilt from the $Z_{15}$ anomaly using equation (1) and transfer function $w(x)$ given by equations (5) and (6). To rebuild the $N_{34}$, we have truncated the sum (5) to the first four EOFs, which cumulatively explain 84% of the overall $Z_{15}$ data set variance. The two dashed lines represent the estimates of the transfer function obtained from the EOF analysis of the $Z_{15}$ anomaly performed using data from the entire observation period (i.e. 1960-2020, red dashed line) and the sub-period 1960-1990 (green dashed line). In both cases, the $N_{34}$ is accurately reconstructed, indicating the stability of the method in relation to the period over which the EOF analysis is performed. This is further supported by the similarity between the transfer functions estimated from the EOF analysis of the $Z_{15}$ anomaly over the entire observation period and the period 1960-1990 (see Figures 1b and 1c). A general analysis of the transfer function's spatial structure indicates that it is characterized by high positive values in the eastern part of the basin, high negative values in the western part, and values close to zero in the central part of the basin. This means that even small variations in the depth of the $Z_{15}$ anomaly in the western or eastern part of the tropical Pacific can have significant impacts on the SST anomaly in the central part of the basin, while even large variations in $Z_{15}$ in the central Pacific have minimal influence on the SST variation in this region.

## 4. Wind stress and $Z_{15}$: "quasi-balance" and residual mean dynamics

We start with the full double layer, linearized shallow water equations, in which the term $\tau/(g'\cdot H\cdot\rho)=\tau/(c_1^2\cdot\rho)$ is expressed as the sum of a solenoidal (i.e. divergence-free) and irrotational (curl-free) vector field, as prescribed by the Helmotz decomposition (see e.g Morse and Feshbach, 1953), i.e.

$$\frac{\tau}{g'\cdot H\cdot\rho}=\frac{\tau}{c_1^2\cdot\rho}=\nabla\phi+\nabla\times\boldsymbol{A} \tag{7}$$

In equation (7), $\tau$ represents the wind stress, $g'$ the reduced gravity, $H$ and $\rho$ the thickness and density of the surface layer, and $c_1^2 = g'H$ the velocity of the first baroclinic mode. The terms $\phi$ and $\boldsymbol{A}$ are commonly referred to as the scalar and vector potential, respectively. Note that the vector $\tau/(c_1^2\cdot\rho)$ is dimensionless, and will be referred to as DWS in the following. Note also that $\phi$ and $\boldsymbol{A}$ have dimensions of length. Furthermore, note that since the DWS lies in the x-y plane, $\boldsymbol{A}=(0,0,S)$. The function $S$, is typically referred to as the DWS stream-function.

Using the Helmotz decomposition of the DWS and the definitions above, the linear shallow water equation can be written in terms of $\phi$ and $S$ as

$$\begin{cases} \frac{1}{g'}\left(\frac{\partial u}{\partial t}-fv\right)=\frac{\partial h}{\partial x}+\frac{\partial \phi}{\partial x}+\frac{\partial S}{\partial y} \\ \frac{1}{g'}\left(\frac{\partial v}{\partial t}+fu\right)=\frac{\partial h}{\partial y}+\frac{\partial \phi}{\partial y}-\frac{\partial S}{\partial x} \\ -\frac{\partial h}{\partial t}+D\left(\frac{\partial u}{\partial x}+\frac{\partial v}{\partial y}\right)=0 \end{cases} \tag{8}$$

where $D$ is a reference depth set to 120 m, $h$ is the departure of the interface layer from $D$ (i.e. $H=D-h$) and the continuity equation has been linearized around $D$ as in Jin (1997). Jin (1997) sought for solutions of the quasi-geostrophic equations, in which $h$ is in equilibrium with the zonal wind stress, and assumed the meridional wind stress to be zero. Here we seek for solutions of (8) around $h_{eq}$, with $h_{eq}$ being the shape of the interface layer at the hydrostatic state (i.e. the shape of the interface layer in equilibrium with the DWS along the zonal and meridional direction), namely $h(x,y,t)=h_{eq}(x,y,t)-q(x,y,t)$. Obviously, $-\nabla^2 h_{eq}=\nabla^2\phi$. This means that the divergence of the DWS, forces the "quasi-equilibrium" part of the dynamics and $q$ describes the deviations of the dynamics from the "quasi-equilibrium". It is therefore natural to set $-h_{eq} = \phi$.

In Figure 2a, the meridional mean of the $h_{eq}$ anomaly is shown in relation to El Niño (indicated by the dots on the right vertical axis) and La Niña (indicated by the squares on the left vertical axis) events. It is evident that the deformation of the interface layer induced by $h_{eq}$ is consistent with both El Niño and La Niña events.

In Figure 2b, we present the $N_{34}$ obtained from equation (1) using the $h_{eq}$ anomaly instead of the $Z_{15}$ anomaly and the transfer function of Figure 1a. Although the correlation coefficient between the measured and reconstructed $N_{34}$ is relatively high (0.83), there are significant differences between the simulated and measured values of the $N_{34}$ during extreme El Niño events. This suggests that while the equilibrium dynamic accurately captures the interannual variability of SST anomalies in the central tropical Pacific, it only partially resolves the fluctuations of the $N_{34}$ over the longer time-scales (i.e. interdecadal or longer). This leads us to hypothesize that $q$, which is forced by the rotational part of the DWS, encapsulates the long-term variability of the $Z_{15}$ anomaly and, thus, of the $N_{34}$. From equation (8), it is possible to derive an equation for $q$. However, a great simplification can be made by using the geostrophic approximation. A reasonable strategy to use this approximation is to exclude the equatorial belt between 0.5°S- 0.5°N and derive an equation for $q$ in the two strips between 0.5°N-5°N and 0.5°S-5°S. The geostrophic approximation appears reasonable because we are interested in the long-term, basin-scale variability of $q$. Using this approximation we find

$$\begin{cases} v = -\left(\frac{g'}{f}\right) \cdot \left(\frac{\partial S}{\partial y} - \frac{\partial q}{\partial x}\right) \\ u = -\left(\frac{g'}{f}\right) \cdot \left(\frac{\partial S}{\partial x} + \frac{\partial q}{\partial y}\right) \end{cases} \tag{9}$$

and, observing that $\partial h_{eq}/\partial t \approx 0$, substituting equation (9) in the continuity equation, we get

$$\frac{\partial q}{\partial t} - D\left(\frac{g' \cdot \beta}{f^2}\right) \cdot \frac{\partial q}{\partial x} = D\left(\frac{g'}{f}\right) \nabla^2 S - D\left(\frac{g' \cdot \beta}{f^2}\right) \frac{\partial S}{\partial y} \tag{10}$$

where $\beta$ is the "equatorial beta-plane" and we have set $f = \beta y$.

To account for the slowly changing background mean climate conditions of the equatorial Pacific, we have considered $g' = g'(t)$, with $g'(t)$ representing the average spatial reduced gravity field over the region 5°S-5°N, 160°E-90°W. This was computed from TS data, as discussed in Appendix B. Figure 3 shows $g'(t)$ and illustrates that the reduced gravity field in the entire tropical Pacific is characterized by a linear trend of increase of approximately $0.66 \cdot 10^{-4}$ m/s² per year, in addition to oscillations over the sub-decadal time-scale. This increase in reduced gravity is a result of an increase in the stability of the water column, which is consistent with an increase of the stratification of the water column due to the uneven increase in temperature of the surface and bottom layers (An et al., 2008).

With the above observation, equation (10) describes an internal wave ($q$-wave in the following) forced by the rotational component of the DWS that propagates eastward with a velocity $c = D \cdot (g' \cdot \beta / f^2) = D \cdot (g'/\beta y^2)$ that depends on time and latitude. With a value of $g' = 2.8 \cdot 10^{-2}$ m/s$^2$, the $q$-wave velocity ranges from 0.48 m/s at 5ºS (or 5ºN) to 48 m/s at 1ºS (or 1ºN). This indicates that $q$-waves are highly dispersive and significantly faster than the first baroclinic mode near the equator.

Figure 4a displays the time-averaged spatial distribution of the DWS streamline function ($S$) anomaly, while Figures 4b and 4c depict the temporal variability of the space-averaged $S$ and $S$ anomaly, respectively. The spatial distribution of $S$ is characterized by a high-value region south of the equator in the eastern part of the basin and a low-value region north of the equator. The temporal variability of $S$ is primarily influenced by the annual signal. Notably, the temporal variability of the spatially-averaged $S$ anomaly peaks during extreme El Niño events in 1982-83, 1997-98, and 2015-16. However, these peaks appear to be a response from the atmosphere rather than a forcing factor.

Equation (10) was solved using the technique outlined in Appendix C. To test the sensitivity of the solution to initial conditions, we solved equation (10) with various initial conditions. In Figure 5a, the time-averaged $q$-wave obtained from an initial condition of zero is shown. Figures 5b and 5c display the time variability of the spatial average of the $q$-wave and the $q$-wave anomaly. Figure 6a shows the time-averaged $q$-wave obtained from an initial condition consisting of a Gaussian with an amplitude of 10 m, an e-folding length of 500 km in the zonal direction, and 350 km in the meridional direction, centered over the equator at the dateline. Figures 6b and 6c show the time variability of the spatial average of the q-wave and the q-wave anomaly obtained from the Gaussian initial condition. Although the spatial patterns displayed in Figures 5a and 6a are slightly different, Figures 5b, 5c, 6b, and 6c reveal that the temporal variability of the average spatial $q$-wave and its anomaly are nearly unaffected by the initial condition and dominated by an interdecadal signal with period of approximately 25-27 years.

To analyze the effect of $q$-wave interdecadal variability on SST anomalies in the central tropical Pacific, we reconstructed the $N_{34}$ using equation (1) and replacing the $Z_{15}$ anomaly with the $h_{eq}$-$q$ anomaly, with $q$ computed with the initial condition set to zero (Figure 5). The results are shown in Figure 7, where we have plotted, beside the estimate of $N_{34}$ obtained by replacing the $Z_{15}$ anomaly with $h_{eq}$-$q$ anomaly, the measured $N_{34}$ (blue line) and the estimated $N_{34}$ using the $h_{eq}$ anomaly (green line, same as Figure 2a) for comparison. It is evident that the inclusion of the $q$-wave significantly improves the accuracy of the $N_{34}$ estimate. In particular, the addition of the $q$-wave allows for a more precise representation of extreme El Niño events, compared to using only $h_{eq}$.

To demonstrate that the improvement in the $N_{34}$ estimate is a result of low frequency signals encapsulated in $q$, which are filtered out by the "quasi-equilibrium" assumption, in Figure 7 we have also plotted the $N_{34}$ estimates obtained by replacing the $Z_{15}$ anomaly with the residual signal of the high-passed $h_{eq}$ minus the low-passed $q$ (represented by a black dashed line) and the $h_{eq}$ minus the high-passed $q$ (represented by a pale green dashed line). The high and low pass filtering was performed with a frequency cut of $10^{-1}$ years$^{-1}$. It can be observed that while the low-pass filtering of $q$ and high-pass filtering of $h_{eq}$ do not significantly alter the $N_{34}$ estimate, except for some short and localized periods, the high-pass filtering of $q$ results in a $N_{34}$ estimate that is similar to that obtained using only $h_{eq}$ (compare the pale green dashed line and green solid line). This demonstrates that the interannual variability of the SST anomaly in the central tropical Pacific can be explained by the equilibrium dynamics, which encapsulates the variability of the interface layer over time scales shorter than 10 years, and is associated with the divergence of the DWS. Conversely, the residual non-equilibrium dynamics, which is forced by the rotational component of the DWS, determines the variability of the SST anomaly over the central tropical Pacific over time-scales longer than 10 year and is responsible for extreme El Niño events.

## 5. Discussion and Conclusion

In this study, we have investigated the relationships between the thermocline depth anomaly, and wind stress in the entire tropical Pacific, and changes in the average SST anomaly in the central part of the basin over both interannual and interdecadal time scales. Our results indicate that the variability in the SST anomaly in the central tropical Pacific can be accurately represented by an integral relationship, which links the variability in the thermocline depth anomaly to the Niño-3.4 index. This integral relationship is defined by a transfer function derived from an EOF analysis of the thermocline depth anomaly, which we have assumed to be represented by the 15º isotherm depth anomaly. The spatial structure of the transfer function indicates that even small variations in the thermocline depth anomaly in the western or eastern parts of the tropical Pacific can have significant impacts on the SST anomaly in the central part of the basin, while large variations in the thermocline depth anomaly in the central Pacific have minimal influence on SST variation in this region. This result is consistent with Zhong et al. (2025), who demonstrated, using a different approach, that El Niño is more predictable in the eastern and western parts of the tropical Pacific compared to the central part of the basin.

Our analysis has shown that the transfer function's structure remains stable regardless of the chosen period for EOF analysis of the thermocline depth anomaly. However, the effectiveness of the transfer function in capturing the interdecadal variability of the $N_{34}$ may vary depending on the specific period selected for the EOF analysis. To definitively address this issue, a systematic sensitivity test would be necessary. We have only performed a few spread tests, which suggest that in order to accurately capture the long-term variability of the $N_{34}$ in relation to changes in the thermocline depth anomaly, the EOF analysis of the thermocline depth anomaly must be conducted over a sufficiently long period to include a range of El Niño and La Niña events with varying intensities. This is evident in the period 1960-1990, which includes 11 El Niño events (four weak, three moderate, three strong, and one extreme) and 14 La Niña events (eight weak, one moderate, and five strong).

To analyze the impact of wind stress variability on SST anomalies in the central tropical Pacific, we examined the coupling between changes in the depth of the thermocline and the irrotational (curl-free) and solenoidal (divergence-free) components of the DWS. The use of the DWS is necessary because, even a wind stress without divergence or curl, can provide divergence or vorticity to the horizontal ocean dynamics through spatial changes of the first baroclinic mode velocity and/or surface density. Our findings indicate that the irrotational component of the DWS, which is related to the divergence of the DWS and drives the quasi-equilibrium dynamics, is the main factor influencing the interannual variability of SST anomalies in the central tropical Pacific. This is consistent with the findings of Eusebi Borzelli and Carniel (2023), who observed that the region of maximum wind convergence in the tropical Pacific shifts along the zonal direction in phase with the SOI. This suggests that the divergence of the wind field can explain the variability of the tropical Pacific climate system on the ENSO time-scale.

In this research, we have demonstrated that the equilibrium dynamics can be described by an equation between the DWS scalar potential and the deformation of the interface layer. This equation takes the form $-\nabla^2 h_{eq} = \nabla^2 \phi$ and we have set $-h_{eq} = \phi$. Although the most natural, this solution is not the most general and this choice could be questioned. A more general solution could be, for instance, $-h_{eq} = \phi + \alpha \cdot x + \beta \cdot y$, where the linear components represent the average slope over the entire observation period of the thermocline in the zonal and meridional directions. However, this signal does not contribute to the $h_{eq}$ anomaly, and for the sake of simplifying the formalism, we have chosen to set $-h_{eq} = \phi$.

Non-equilibrium dynamics, driven by the solenoidal component of the DWS, which encapsulates the rotational part of the DWS, plays a significant role over time scales longer than a decade. This rotational component is responsible for the generation of internal waves, referred to as "$q$-waves", which propagate eastward and deform the thermocline. These $q$-waves contribute to changes in the SST anomaly in the central tropical Pacific. Therefore, $q$-waves represent the impact of non-equilibrium dynamics on changes in the SST anomaly in this region. While the rotational component of the DWS does not exhibit variability over time scales longer than a decade, it acts on a stratified ocean. The stratification regime of this ocean, as indicated by the reduced gravity field, changes over various time scales, including the interdecadal. Our findings suggest that the interdecadal variability of the SST anomaly in the central tropical Pacific is a result of the interplay between the variability of the reduced gravity and the rotational component of the DWS.

The uneven heating of the surface and bottom layers of the tropical Pacific under greenhouse warming described by An et al. (2008) has the obvious consequence of increasing the stability of the water column and, therefore, increasing the reduced gravity. However, the role of surface salinity in determining changes in the reduced gravity field is less certain, although some authors have emphasized the importance of including sea surface salinity records to improve El Niño forecasting (e.g. Wang et al., 2024). In addition to the redistribution of salt by ocean currents, the acceleration of the hydrological cycle under greenhouse warming also contributes to regional means and long-term trends in sea surface salinity (Durack et al., 2012). However, it has been shown that estimates of long-term trends in sea surface salinity in the tropical Pacific are regionally dependent (Zhi et al., 2025). While our research did not specifically address the impact of long-term surface salinity trends on the variability of the reduced gravity field, a focused analysis using the methods presented here could provide insight into the role of salinity changes in the interdecadal variability of SST anomalies in the tropical Pacific.

As a final remark, some authors have noted that the 2023-24 El Niño event was unusual due to its significant oceanic warming, but relatively subdued Southern Oscillation and wind anomalies over the tropical Pacific (Pang et al., 2025). This suggests that the event was not influenced by the traditional Bjerknes feedback mechanism. According to these authors, the intense oceanic warming was mainly caused by the accumulation of heat content in the western Pacific during the preceding prolonged La Niña period, and they attributed the event to oceanic processes. In the terminology of this research, an increased heat piled up in the western tropical Pacific is expected to contribute to increasing the stability of the water column and, in turn, increase the reduced gravity. According to Pang et al. (2025), the 2023-24 El Niño was dominated by interannual variability and the geostrophic approximation used here is only able to capture variability over a time-scale longer than 3-4 years. However, it is interesting to note that, according to equation (10), an increased reduced gravity determines a more efficient forcing of q-waves by the rotational component of the dimensionless wind stress. In order to clarify this issue, more specific approximations that allow for the derivation of an equation capable of capturing the interannual variability of q-waves, based on, for instance, the quasi-geostrophic approximation used in Jin (1997), are necessary.

**Appendix A: solution of equation (3) and derivation of equation (5)**

We have used dimensionless EOF analysis of the $Z_{15}$ anomaly. This EOF algorithm is summarized by the following equations

$$\begin{aligned}\mathbf{Z}(t) &= \sum_{m=1}^{M} \sqrt{\lambda_m}\, \alpha_m(t) \cdot \mathbf{Y}_m(x_i) \\ \alpha_n(t_j) &= \frac{1}{\sqrt{\lambda_n}} [\mathbf{Z}(t)]^T \cdot \mathbf{Y}_n(x_m)\end{aligned} \qquad (A.1)$$

that, for convenience we have rewritten in vector form defining the column vectors $\mathbf{Z}(t)$, whose $i^{th}$ element is $Z(x_i,t)$ $(i=1...M)$, and $\mathbf{Y}_m$, whose $i^{th}$ element is the value of the $m^{th}$ $(m=1...M)$ EOF at point $x_i$. In equation (A.1), the suffix $T$ indicates the transpose. After rewriting equation (3) in matrix form, observing that $\mathbf{Y}_k$ are eigenvectors of $\mathbf{A}$, with an obvious notation, expressing $\mathbf{w}=q_1\mathbf{Y}_1+...+q_M\mathbf{Y}_M$ and $\mathbf{n}=b_1\mathbf{Y}_1+...+b_M\mathbf{Y}_M$ we have $q_k = b_k/\lambda_k$, where $b_k$ is the projection of $\mathbf{n}$ along the direction of the $k^{th}$ EOF, i.e. $b_k = \mathbf{n}^T \cdot \mathbf{Y}_k$. Therefore, $\quad b_k = \mathbf{n}^T \cdot \mathbf{Y}_k = \int_0^T \frac{dt}{T} N_{34}(t) \cdot [\mathbf{Z}(t)]^T \cdot \mathbf{Y}_k = \sqrt{\lambda_k} \int_0^T \frac{dt}{T} N_{34}(t) \cdot \alpha_k(t) \quad$ or

$$\mathbf{w} = \sum_{n=1}^{M} \frac{\mathbf{Y}_n}{\sqrt{\lambda_n}} \cdot \int_0^T \frac{dt}{T} N_{34}(t) \cdot \alpha_n(t) \qquad (A.2)$$

which is equation (5).

**Appendix B: determination of the reduced gravity field and first baroclinic mode velocity**

To compute the reduced gravity and the velocity of the first baroclinic mode we used the strategy suggested by Eusebi Borzelli and Carniel (2023) and Eusebi Borzelli et al. (2024). Firstly, the ocean's vertical structure was assumed to consist of two layers, and the density profiles were fitted with a step function. This involved dividing each profile into two sub-segments of varying length and calculating the average density for each segment. The resulting step function representations were then compared to the original profile, and the one with the lowest root-mean-square was selected as the best step function representation of the water column. This process was repeated for each date of the observation period, providing spatial distributions of density in the surface ($\rho(x,y,t)$) and bottom ($r(x,y,t)$) layers and a "first approximation" depth of the surface layer ($H^0(x,y,t)$). In the calculation of the velocity of the first baroclinic mode, we excluded all points in which $H^0(x,y,t)$ was less deep than 60 m (with 60 m supposed to be the maximum depth of the mixed layer) and $\rho(x,y,t) \geq r(x,y,t)$. This was done to avoid spurious results due to density changes in the mixed layer and instabilities in the water column. The velocity of the first baroclinic mode was obtained by solving the vertical mode equation (Wusch and Stammer, 1997), namely

$$\frac{d}{dz}\left[\frac{1}{N^2(x,y,z,t)} \cdot \frac{dP_n}{dz}\right] + \frac{1}{c^2_n(x,y,t)} P_n = 0 \tag{B.1}$$

subject to rigid upper and lower boundary conditions, at each date of the observation period, in each point of the observation domain. In equation (B.1), $N^2(x,y,z,t)$ is the buoyancy frequency, $n=0,1...$, with $n=0$ indicating the barotropic mode, and $1/c^2_n$ are real-valued, positive eigenvalues, with $c_n$ indicating the velocity of the $n^{th}$ baroclinic mode. In the final calculation of the depth of the surface layer, we excluded all points in which $H^0(x,y,t)$ was less deep than 60 m (with 60 m supposed to be the maximum depth of the mixed layer) and $\rho(x,y,t) \geq r(x,y,t)$. This was done to avoid spurious results due to density changes in the mixed layer and instabilities in the water column.

**Appendix C: Solution of equation (10).**

Equation (1) was transformed in to a system of ordinary independent differential equations using the Fourier method. Observing that in equation (10) the latitude is a parameter, at every latitude it can be rewritten as

$$\frac{\partial q}{\partial t} - c_y(t) \cdot \frac{\partial q}{\partial x} = b_y(x,t) \quad \text{where} \quad \begin{cases} c_y(t) = D\left(\frac{g' \cdot \beta}{f^2}\right) \\ b_y(x,t) = D\left(\frac{g'}{f}\right) \nabla^2 S - D\left(\frac{g' \cdot \beta}{f^2}\right) \frac{\partial S}{\partial y} \end{cases} \tag{C.1}$$

Fixing the latitude, indicating with $Q(k,t)$ and $B(k,t)$ the Fourier transforms along the zonal directions of $q(x,t)$ and $b_y(x,t)$, we get a system of decoupled ordinary differential equation of the form

$$\frac{dQ}{dt} - ik \cdot c_y(t) Q = B(k,t) \quad \text{with initial condition} \quad Q(k,t=0) = Q_0(k) \tag{C.2}$$

where $Q_0(k)$ is the Fourier transform of the initial condition $q(x,t=0) = q_0(x)$.

We solved equation (C.2) using the integrating factor. Namely, setting $\mu(t) = \exp\left[-ik \int_0^t c_y(s) \cdot ds\right]$ we have

$$Q(k,t) = \frac{Q_0(k) + \int_0^t ds\, B(k,s) \cdot \mu(s)}{\mu(t)} \tag{C.3}$$

**Open Research Section**

Data used in this research can be downloaded succumbing to European Union (EU) regulations on geophysical data exchange (see https://www.copernicus.eu/en/international-cooperation-area-data-exchange).

Original ERA5 wind and SST data can be downloaded from the Copernicus data archive by registered users at

https://cds.climate.copernicus.eu/datasets/reanalysis-era5-single-levels-monthly-means?tab=overview

Original monthly T-S data were made available by the Institute of Atmospheric Physics (IAP) of the Chinese Academy of Sciences (CAS) and can be downloaded the at the following web site

http://www.ocean.iap.ac.cn/pages/dataService/dataService.html?navAnchor=dataService

Matlab scripts, developed for this research under Matlab release 2018a, can be downloaded at
https://doi.org/10.6084/m9.figshare.30609104

**Competing Interests**

The authors declare no competing interest.

**Authors' contributions**

All the authors contributed equally to the work. G.L.E.B. and S.C. proposed the original idea and designed the research. All authors contributed to analyze the data and to interpret the results. G.L.E.B. and C.E.C. provided the figures. G.L.E.B. wrote the initial version of the manuscript, S.C., C.E.C. and M.S. provided corrections and comments.

**Figure captions**

**Figure 1:** a) The blue solid line represents the measured $N_{34}$ values, while the dashed lines represent the simulated $N_{34}$ values from the first six EOFs of the $Z_{15}$ anomaly. The red dashed line shows the $N_{34}$ values simulated from the transfer function calculated using EOF analysis from 1960 to 2020, while the green dashed line shows the $N_{34}$ values simulated from the transfer function calculated from EOF analysis from 1960 to 1990. Colored dots on the bottom horizontal axis represent El Niño events obtained from https://ggweather.com/enso/oni.htm. Color of the dots represent El Niño event intensity, with the following legend: red, extreme; magenta, strong; blue, moderate; green, weak. b) $Z_{15}$ anomaly-$N_{34}$ transfer function calculated from the first six EOFs of the $Z_{15}$ over the period 1960-2020. c) $Z_{15}$ anomaly-$N_{34}$ transfer function calculated from the first six EOFs of the $Z_{15}$ over the period 1960-1990.

**Figure 2:** a) Meridional mean of the dimensionless wind stress potential anomaly. Colored dots on the right vertical axis represent El Niño events classified as discussed in the legend of Figure 1a. Colored rectangles on the left vertical axis represent La Niña events obtained from https://ggweather.com/enso/oni.htm, with the with the following legend: magenta, strong; blue, moderate; green, weak. b) Measured (blue line) and rebuilt $N_{34}$ (green line) obtained with the transfer function of Figure 1b and substituting, in place of the $Z_{15}$ anomaly, the anomaly of $h_{eq}$. Colored dots on the horizontal bottom axis represent El Niño events classified as discussed in the legend of Figure 1a.

**Figure 3:** Spatial average of the reduced gravity field over the region 5ºS-5ºN, 160ºE-90ºW. The blue line represents raw data; the red line represents 5-month median filtered data; the black dashed line is the linear fit.

**Figure 4:** a) Time average of the dimensionless wind stress anomaly. b) Spatial average over the region 5ºS-5ºN, 160ºE-90ºW of the dimensionless wind stress. c) Spatial average over the region 5ºS-5ºN, 160ºE-90ºW of the dimensionless wind stress anomaly. Colored dots on the horizontal top axis of panels b and c represent El Niño events classified as discussed in the legend of Figure 1a.

**Figure 5:** q-waves obtained with initial condition $q_0 = 0$. a) Time average of the deformation of the interface layer deformation caused by q-waves. b) Spatial average of the interface layer deformation caused by q-waves. c) Spatial average of the deformation of the interface layer

deformation anomaly caused by q-waves. In panels b and c, the red dashed line represents low-pass filtered data with cutting frequency $10^{-1}$ years$^{-1}$. The black line represents raw data. Colored dots on the horizontal top axis of panels b and c represent El Niño events classified as discussed in the legend of Figure 1a.

**Figure 6:** Same as Figure 5, but with q-waves obtained from the initial condition $q_0=10 \cdot exp\{-[(x-x_0)/l]^2-(y/m)^2\}$ with $x_0 = 180°$, $l = 500$ km, $m = 350$ km.

**Figure 7:** Measured (blue solid line) and simulated $N_{34}$. Red solid line: $N_{34}$ simulated using the transfer function of Figure 1b and substituting, in place of the $Z_{15}$ anomaly, the anomaly of the interface layer computed as $h = h_{eq} - q$. Green line: $N_{34}$ simulated using the transfer function of Figure 1b and substituting, in place of the $Z_{15}$ anomaly, the anomaly of the interface layer computed as $h = h_{eq}$ (same as green line of Figure 2). Black dashed line: $N_{34}$ obtained from high passed $h_{eq}$ and low passed $q$. Pale green dashed line: $N_{34}$ obtained from $h_{eq}$ and high passed $q$. In high and low pass filtering the cutting frequency is $10^{-1}$ years$^{-1}$. Colored dots on the horizontal bottom axis represent El Niño events classified as discussed in the legend of Figure 1a.

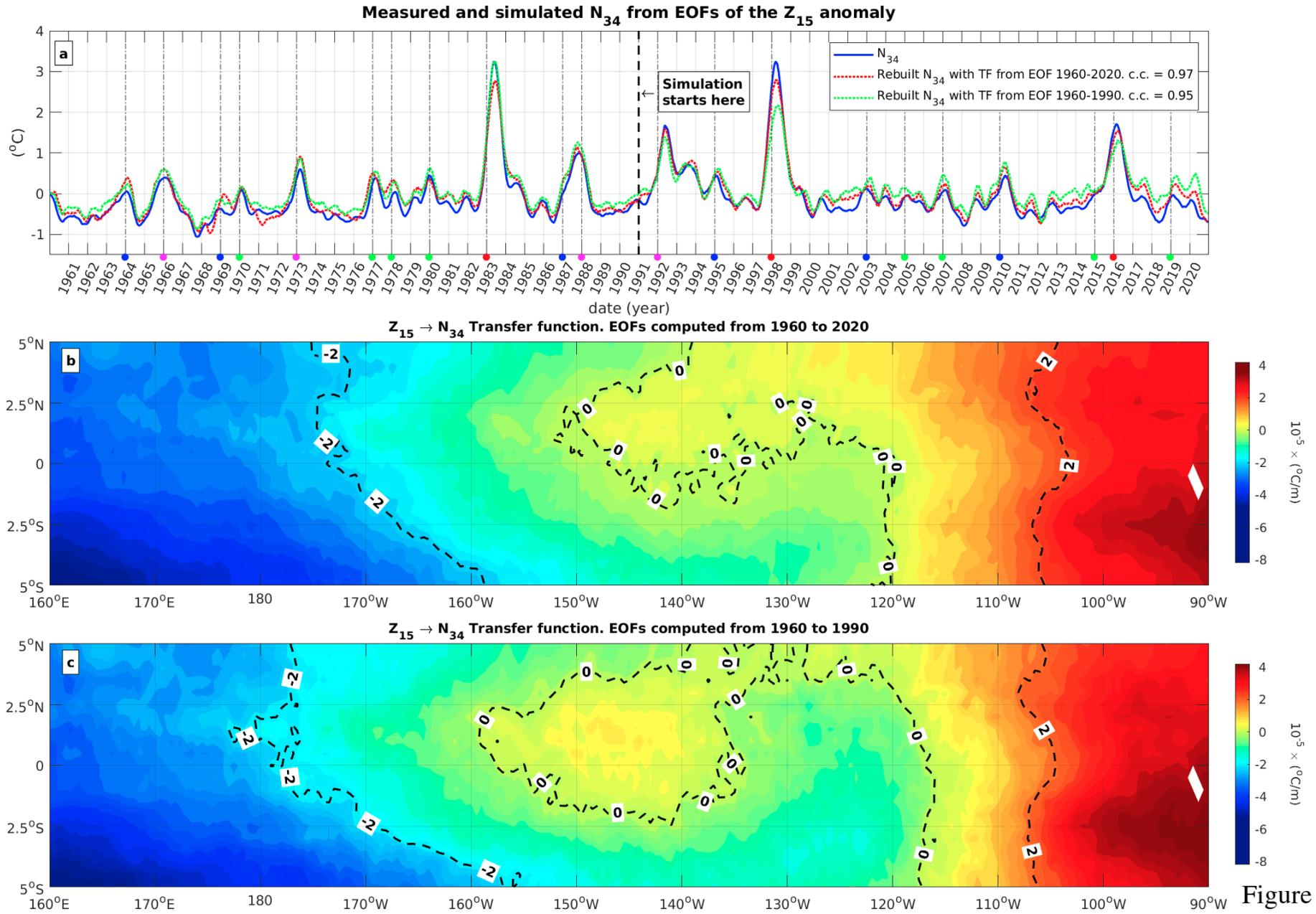

Figure 1

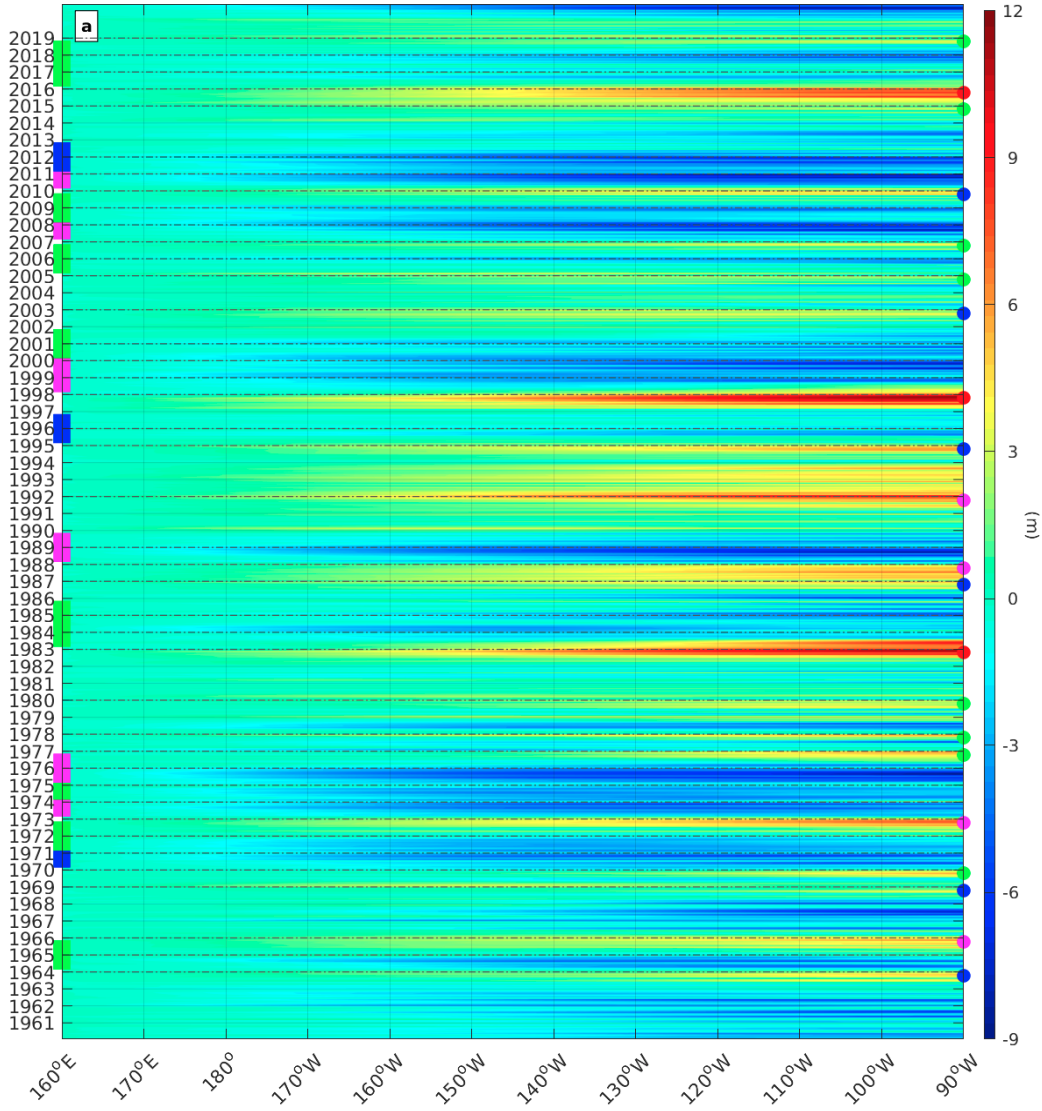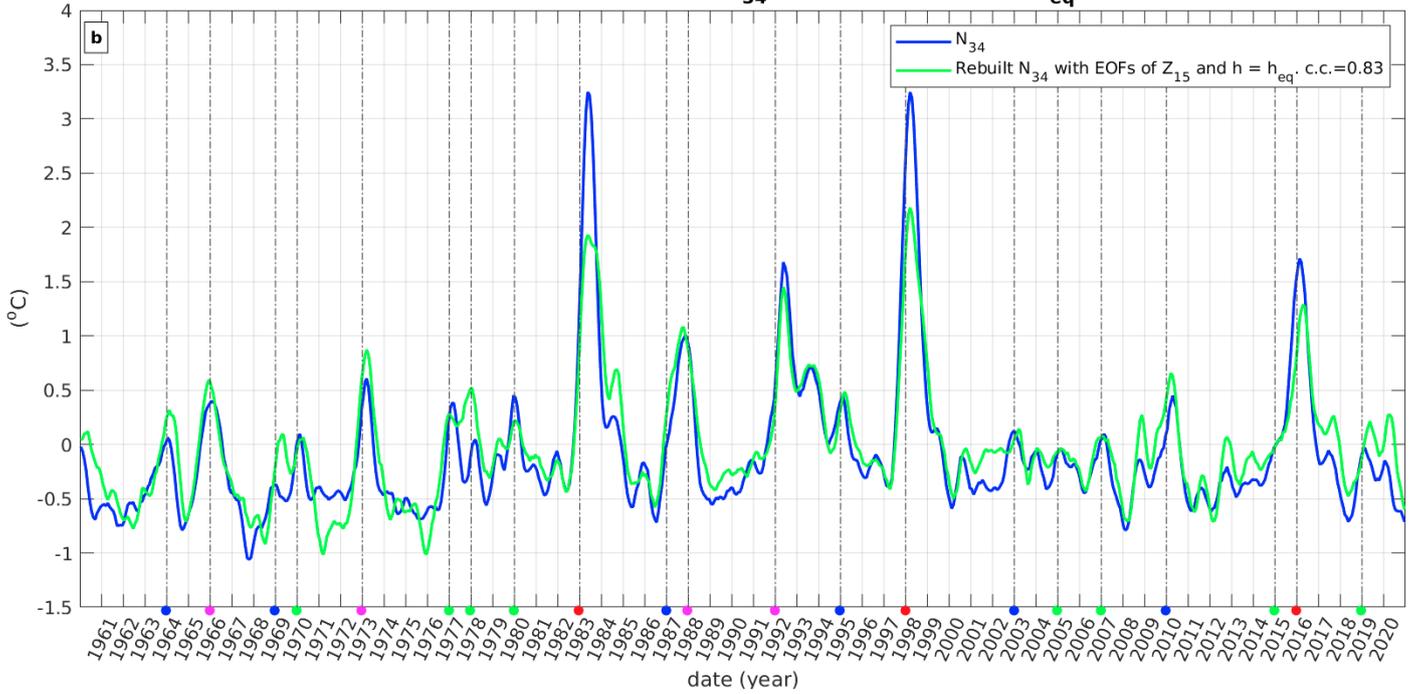

Figure 2

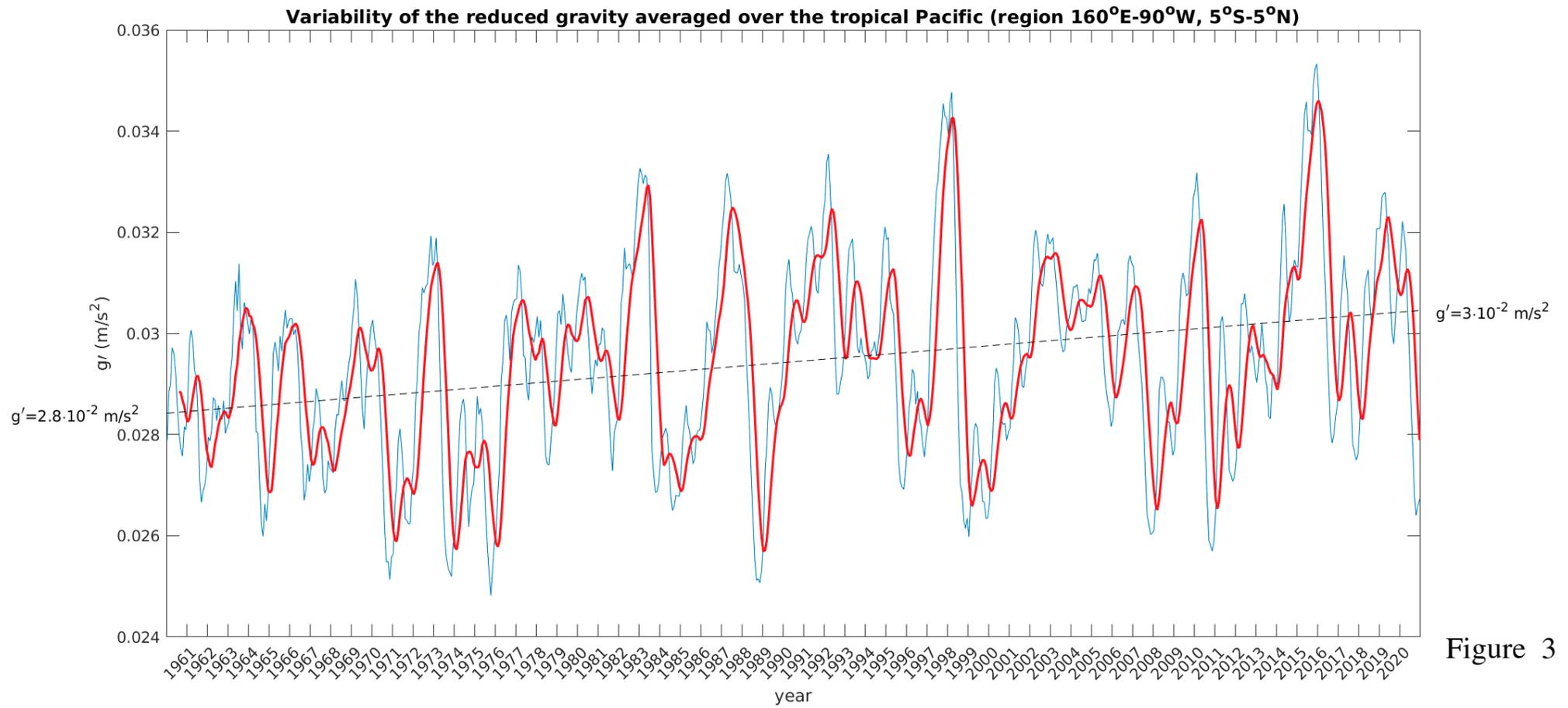

Figure 3

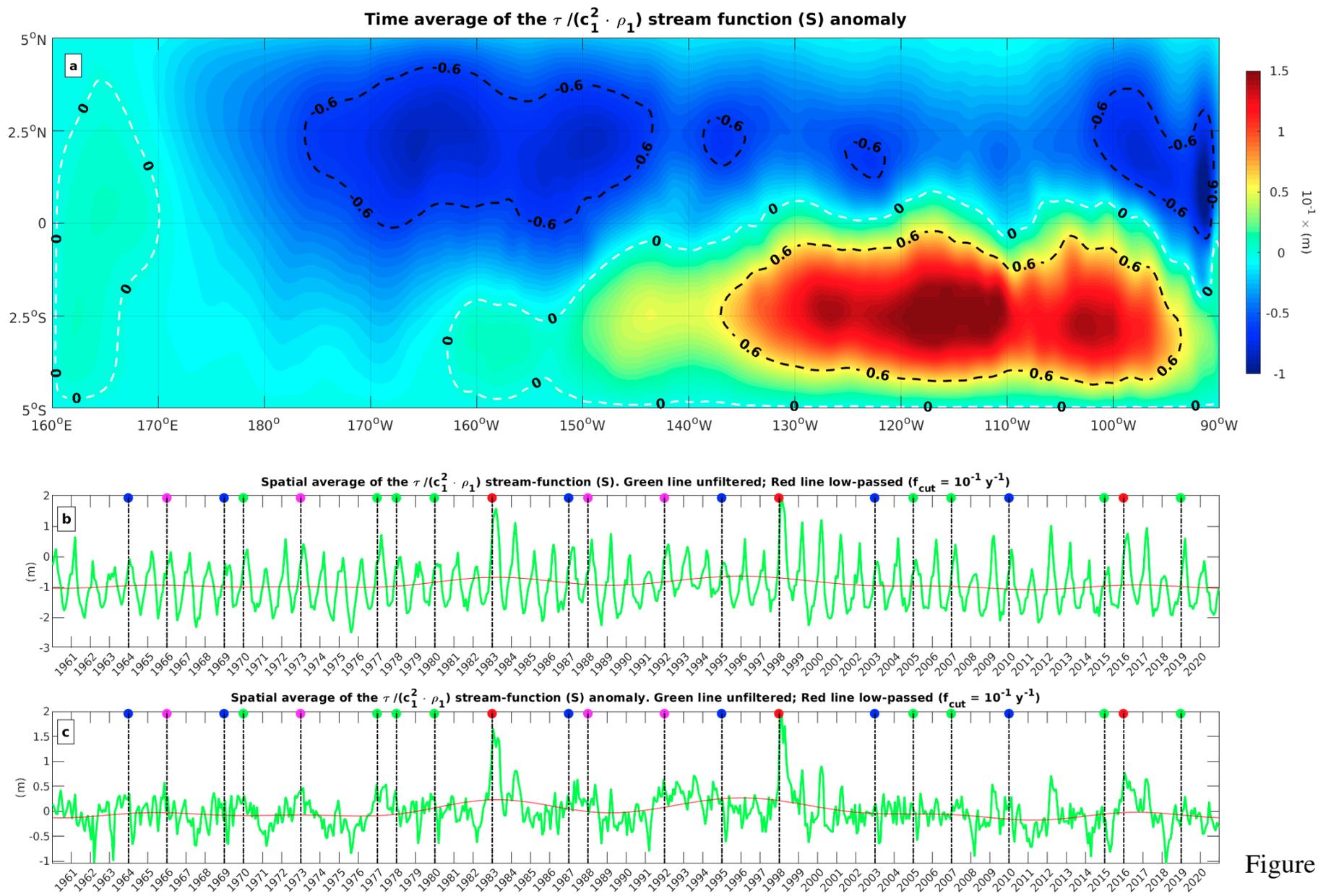

Figure 4

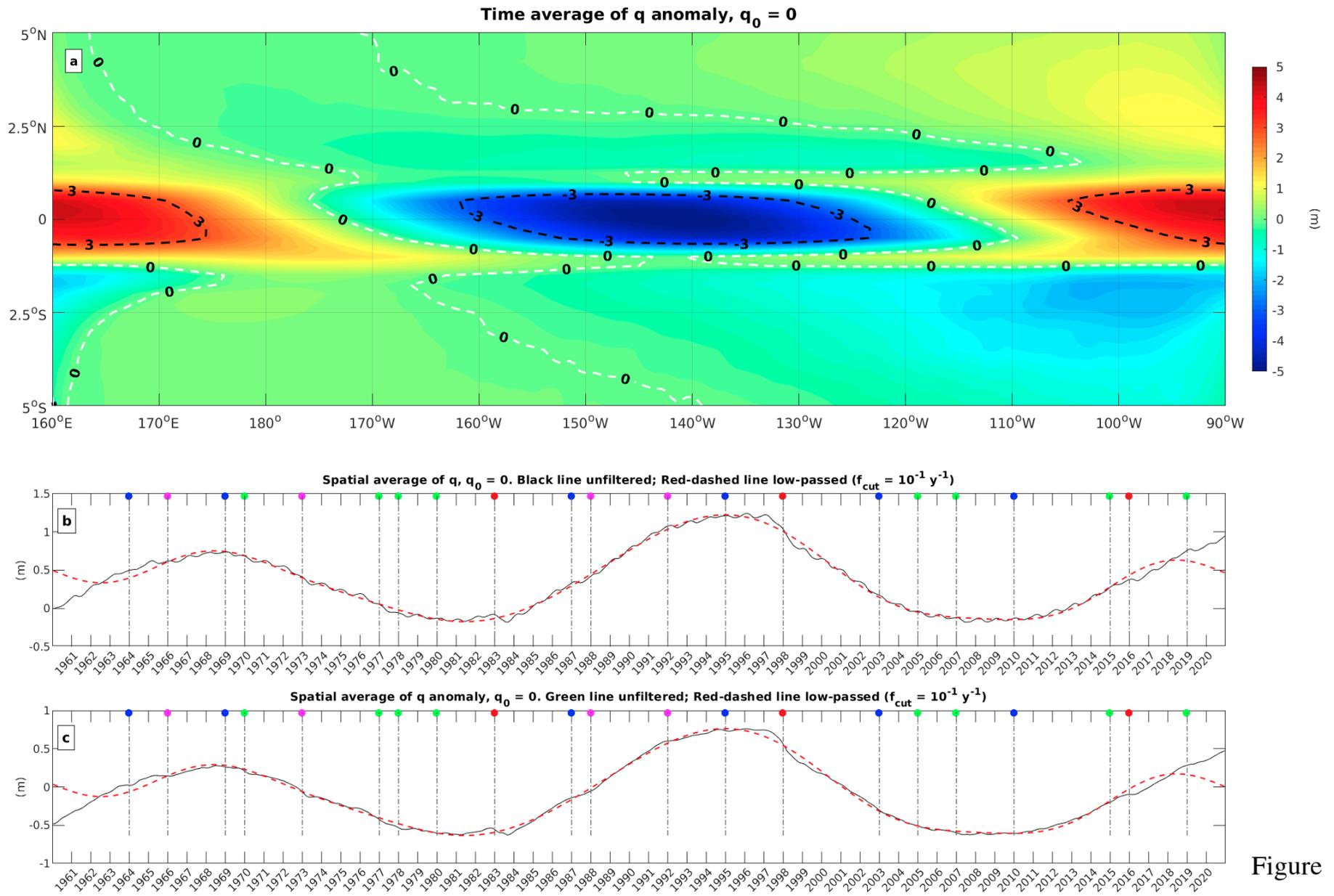

Figure 5

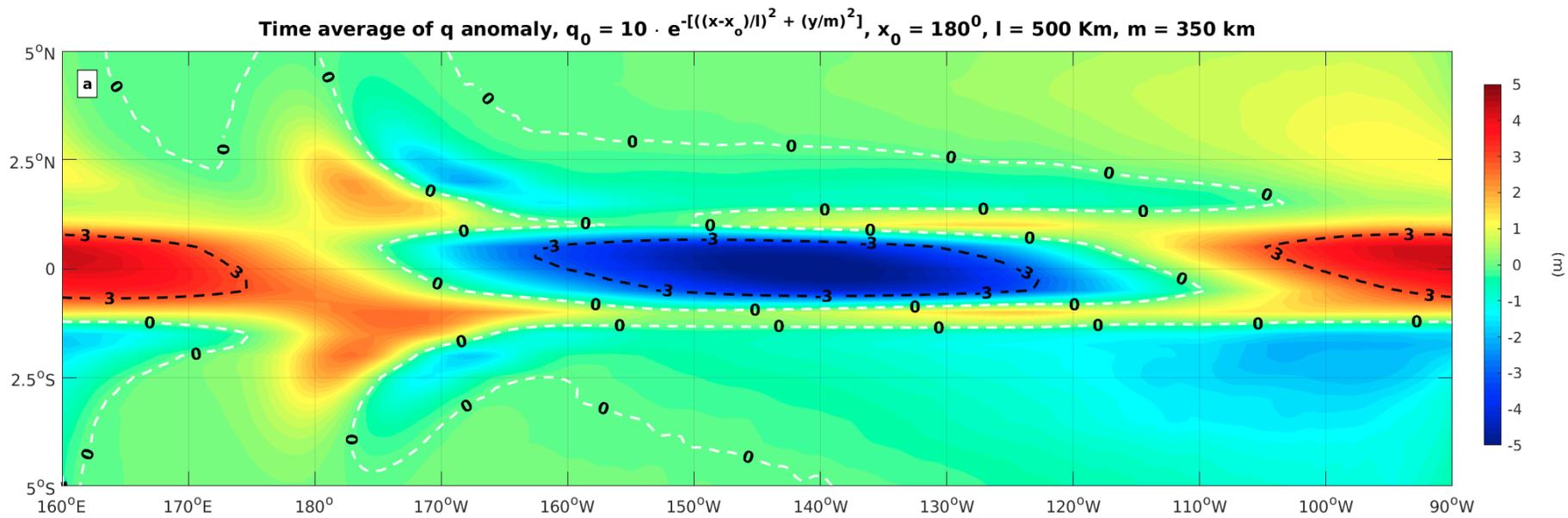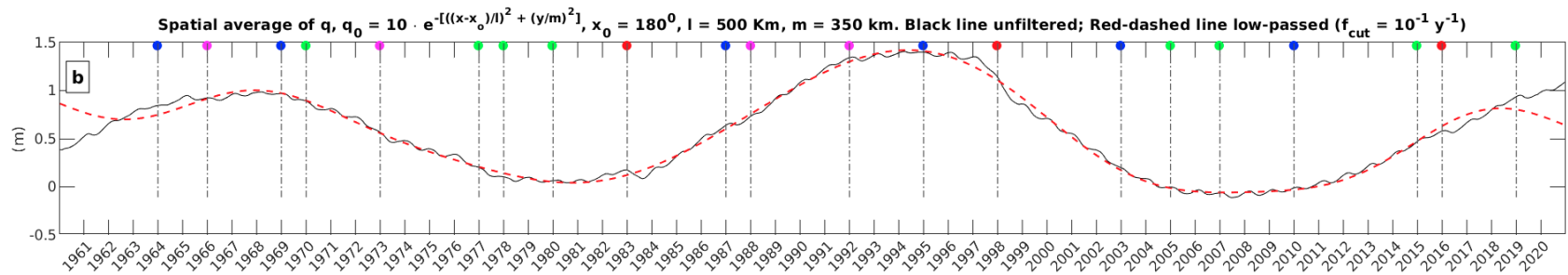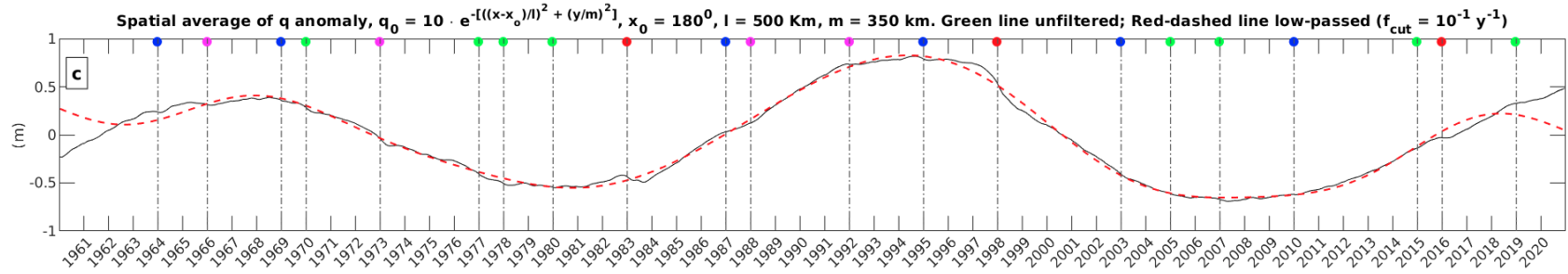

Figure 6

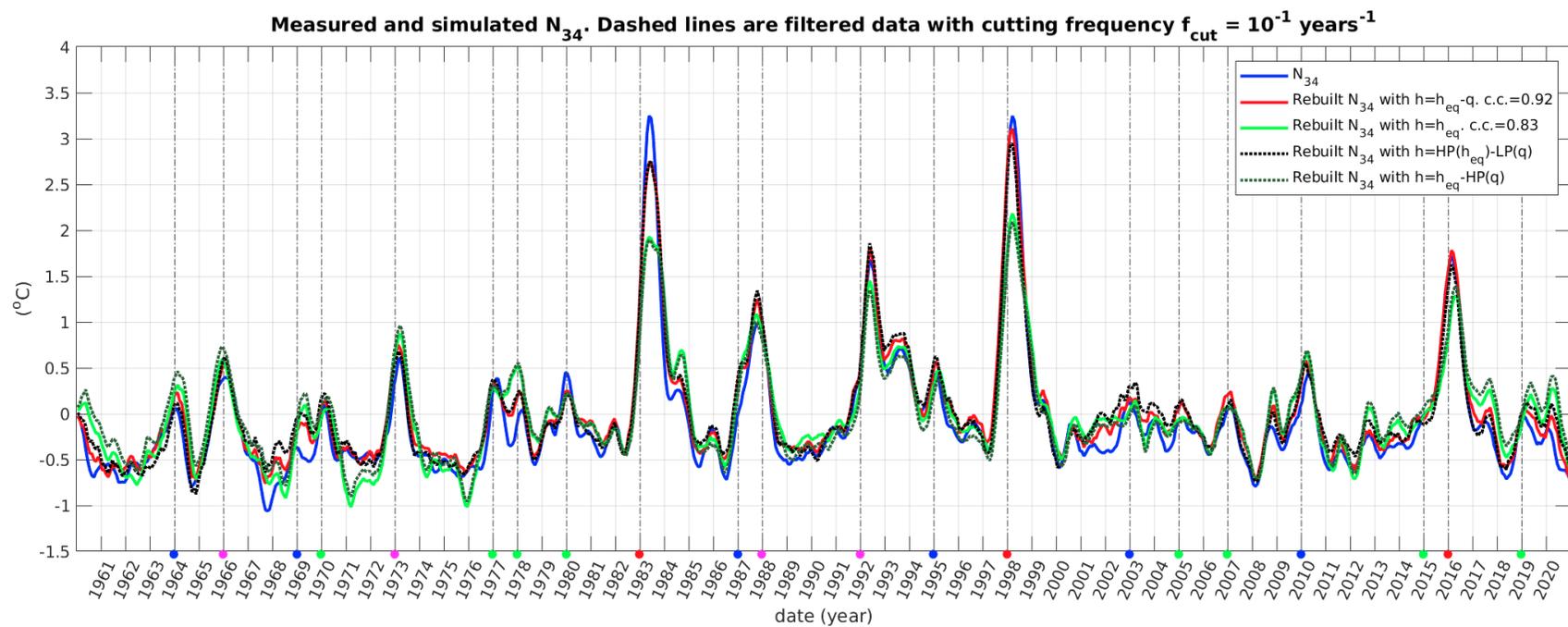

Figure 7